\documentclass{SciPost}

\hypersetup{
    colorlinks,
    linkcolor={red!50!black},
    citecolor={blue!50!black},
    urlcolor={blue!80!black}
}

\usepackage[bitstream-charter]{mathdesign}
\DeclareSymbolFont{usualmathcal}{OMS}{cmsy}{m}{n}
\DeclareSymbolFontAlphabet{\mathcal}{usualmathcal}

\fancypagestyle{SPstyle}{
\fancyhf{}
\lhead{\colorbox{scipostblue}{\bf \color{white} ~SciPost Physics Proceedings }}
\rhead{{\bf \color{scipostdeepblue} ~Submission }}

\fancyfoot[C]{\textbf{\thepage}}
}

\usepackage{wrapfig}
\usepackage[capitalise]{cleveref}

\begin{document}

\pagestyle{SPstyle}

%%% TITLE %%%
\begin{center}{\Large \textbf{\color{scipostdeepblue}{
The Forward Physics Facility at the HL-LHC and\\its Synergies with Astroparticle Physics\\
}}}\end{center}

%%% AUTHORS %%%
\begin{center}\textbf{
Dennis Soldin\textsuperscript{1$\star$}
}\end{center}

%%% AFFILIATIONS / EMAIL %%%
\begin{center}
{\bf 1} Department of Physics and Astronomy, University of Utah,\\
Salt Lake City, UT 84112, USA\\[\baselineskip]

$\star$ \href{mailto:dennis.soldin@utah.edu}{\small dennis.soldin@utah.edu}
\end{center}

%%% CONFERENCE HEADER %%%
\definecolor{palegray}{gray}{0.95}
\begin{center}
\colorbox{palegray}{
  \begin{tabular}{rr}
  \begin{minipage}{0.36\textwidth}
    \includegraphics[width=55mm]{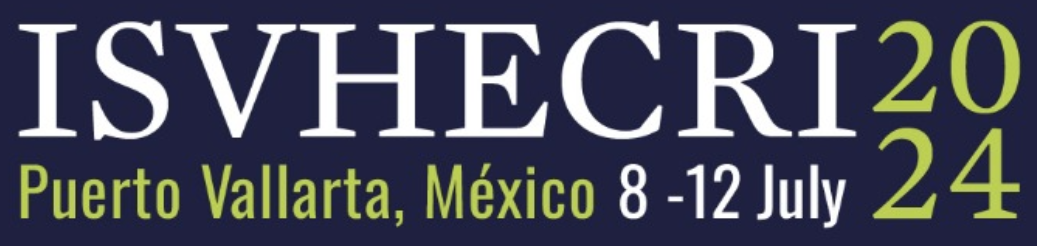}
  \end{minipage}
  &
  \begin{minipage}{0.55\textwidth}
    \begin{center} \hspace{5pt}
    {\it 22nd International Symposium on Very High \\Energy Cosmic Ray Interactions (ISVHECRI 2024)} \\
    {\it Puerto Vallarta, Mexico, 8-12 July 2024} \\
    \doi{10.21468/SciPostPhysProc.?}\\
    \end{center}
  \end{minipage}
\end{tabular}
}
\end{center}

%%% ABSTRACT %%%
\section*{\color{scipostdeepblue}{Abstract}}
\textbf{\boldmath{
High-energy collisions at the high-luminosity Large Hadron Collider (HL-LHC) will generate a vast flux of particles along the beam collision axis, a region not accessible by current LHC experiments. The study of multi-particle production in the far-forward region is especially important for astroparticle physics. High-energy cosmic rays create extensive air showers (EAS) in the atmosphere, driven by hadron-ion collisions in the non-perturbative QCD regime. Therefore, understanding high-energy hadronic interactions in the forward region is crucial for interpreting EAS data and estimating backgrounds for searches of astrophysical neutrinos, among other applications. The Forward Physics Facility (FPF) is a proposal to construct a new underground cavern at the HL-LHC, hosting various far-forward experiments designed to detect particles outside the current LHC acceptance. We will outline the current plans for the FPF and highlight its synergies with astroparticle physics. Specifically, we will discuss how FPF measurements will enhance the modeling of high-energy interactions in the atmosphere, helping to reduce the associated uncertainties in multi-messenger astrophysics.
}}

\vspace{\baselineskip}

%%% COPYRIGHT %%%
\noindent\textcolor{white!90!black}{%
\fbox{\parbox{0.975\linewidth}{%
\textcolor{white!40!black}{\begin{tabular}{lr}%
  \begin{minipage}{0.6\textwidth}%
    {\small Copyright attribution to authors. \newline
    This work is a submission to SciPost Phys. Proc. \newline
    License information to appear upon publication. \newline
    Publication information to appear upon publication.}
  \end{minipage} & \begin{minipage}{0.4\textwidth}
    {\small Received Date \newline Accepted Date \newline Published Date}%
  \end{minipage}
\end{tabular}}
}}
}

\newpage

%%% TOC %%%
%\vspace{10pt}
%\vspace{5pt}
\noindent\rule{\textwidth}{1pt}
\tableofcontents
\noindent\rule{\textwidth}{1pt}
%\vspace{10pt}

%%% SECTION 1 %%%
\section{Introduction}
\label{sec:intro}

Cosmic rays with energies exceeding $10^{11}\,\mathrm{GeV}$ enter Earth's atmosphere, where they interact with air molecules, generating extensive air showers (EASs) that can be detected by large ground-based detector arrays. To infer the properties of the primary cosmic rays, such as their energy and mass, from indirect measurements of secondary particles detected at ground level, simulations are required to interpret the EAS development. A key challenge in understanding EASs is accurately modeling hadronic interactions in the forward region across a broad range of energies, a domain that current collider facilities are unable to directly probe.

\begin{figure}[tb]
    \vspace{-1.em}
    \centering
    \captionsetup{width=1.\linewidth}
    \includegraphics[width=0.9\textwidth]{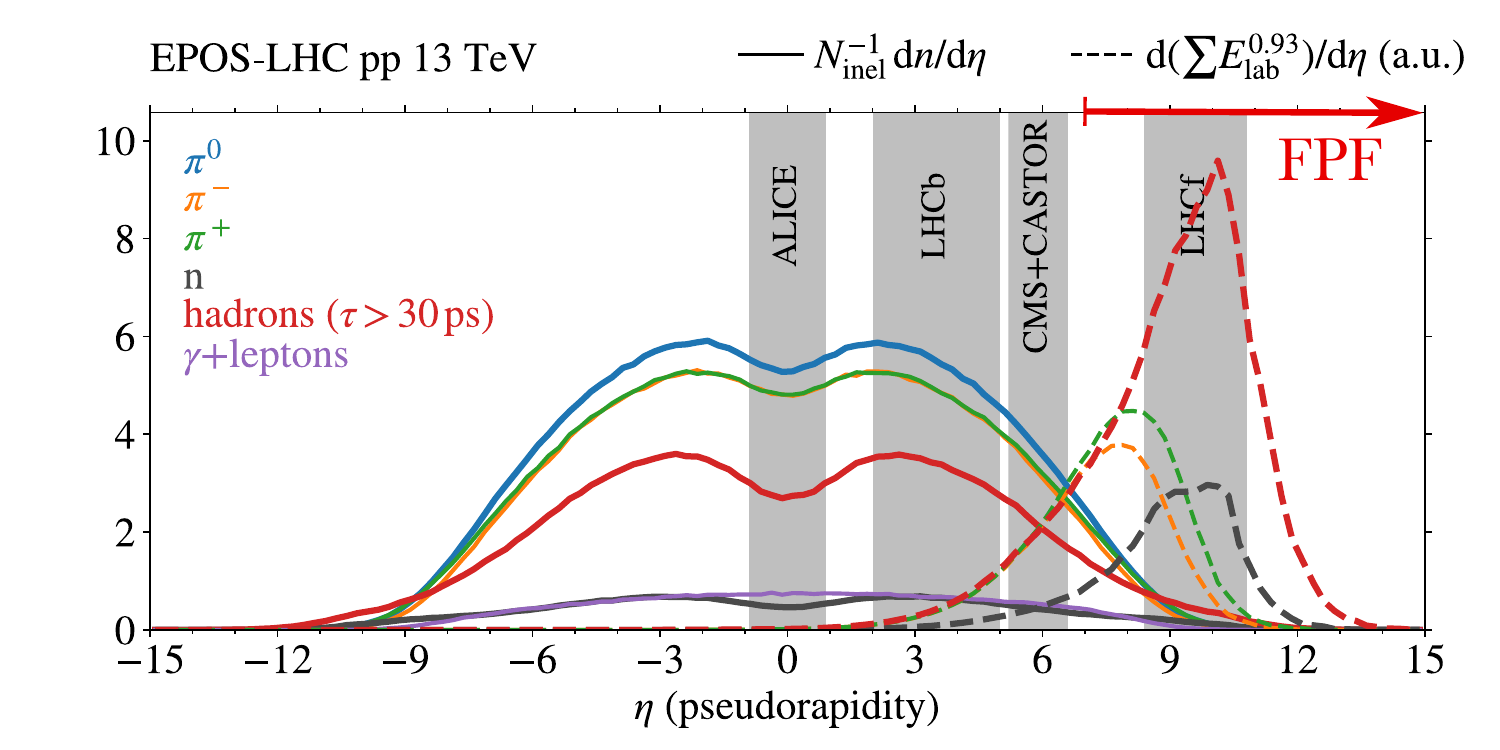}
    \vspace{-.5em}
    
    \caption{\label{fig:eta}Simulated densities of particles~\protect\cite{Albrecht:2021cxw} in arbritrary units (solid lines) in proton-proton collisions using EPOS-LHC~\cite{Pierog:2013ria}. Dashed lines show the estimated number of muons produced by these particles, assuming an equivalent energy for the fixed target collisions in the laboratory frame, $E_{\rm{lab}}$, and $N_\mu \propto E_{\rm{lab}}^{0.93}$.
    }
  %  \vspace{-0.5em}
\end{figure}

\Cref{fig:eta} illustrates simulated particle densities from proton-proton collisions (solid lines), with pseudorapidity (\(\eta\)) ranges relevant to existing LHC experiments overlaid~\cite{Albrecht:2021cxw}. The dashed lines represent the estimated muon densities (\(N_\mu\)), with the assumption \(N_\mu \propto E^{0.93}\), where \(E_{\text{lab}}\) is the laboratory energy of the secondary EAS particles. While mid-rapidity regions have a minimal effect on EAS particle production, the forward region (\(\eta > 4\)) has a profound influence on EAS development, primarily driven by relativistic hadron-ion collisions in the atmosphere at low momentum transfer, within the non-perturbative QCD regime.

Given the difficulty in modeling hadron production from first principles, combined with the lack of sufficient data from current collider experiments, simulations typically rely on phenomenological models of hadronic interactions, which introduce significant uncertainties. As a result, accurate measurements of hadronic interactions, particularly in the far-forward region, are crucial for validating and refining current EAS models.

The Forward Physics Facility (FPF) is a proposal to build a new underground cavern at CERN to house a suite of forward experiments during the high-luminosity LHC (HL-LHC) era~\cite{Anchordoqui:2021ghd, Feng:2022inv, Adhikary:2024nlv}. These experiments will cover the blind spots of the existing LHC detectors and probe high-energy hadronic interactions in the far-forward region. These measurements will improve the modeling of hadronic interactions in the atmosphere, which will help reduce uncertainties in air shower observations and lead to a better understanding of the properties of the highest-energy cosmic rays. Additionally, atmospheric neutrinos produced in EASs in the far-forward region represent a significant background in searches for high-energy astrophysical neutrinos using large-scale neutrino telescopes. As a result, FPF measurements will be critical for refining our understanding of the atmospheric neutrino flux and reducing associated uncertainties in high-energy astrophysical neutrino searches. 

%------------------------
%------------------------

%%% SECTION 2 %%%
\section{The Forward Physics Facility}
\label{sec:facility}

%------------------------
\begin{figure}[tb]
    \centering
    \captionsetup{width=1.\linewidth}
    \vspace{-1em}
    \includegraphics[width=1.\textwidth]{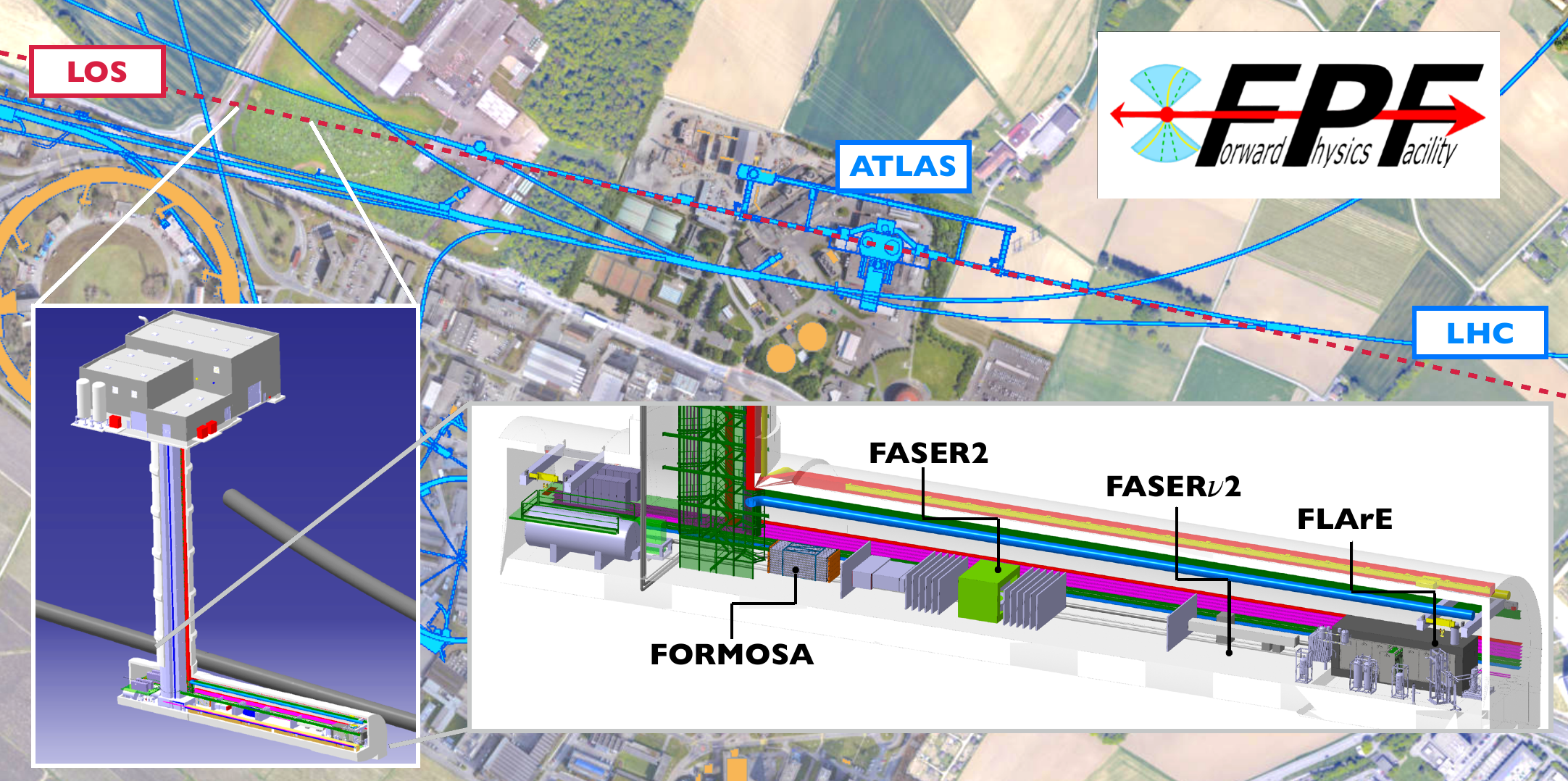}
    \caption{\label{fig:ExecutiveSummaryMap}The FPF is located 627--702\,m west of the ATLAS interaction point along the line-of-sight (LOS)~\cite{Feng:2022inv,Adhikary:2024nlv}. The FPF cavern is 75\,m long and 12\,m wide and will house a diverse set of experiments to fully explore physics in the forward region (see text for details).}
\end{figure}
%------------------------

The Forward Physics Facility~\cite{Anchordoqui:2021ghd, Feng:2022inv, Adhikary:2024nlv} is a proposed underground laboratory at CERN, specifically designed to house a range of experiments focused on the far-forward region at the HL-LHC. Existing LHC detectors leave certain areas along the beam line uninstrumented, potentially missing crucial physics signals from the significant flux of particles produced in the forward direction. In the absence of the FPF, the HL-LHC would lack sensitivity to key phenomena such as neutrino interactions and a variety of particles predicted by extensions to the Standard Model. %, including new force carriers, sterile neutrinos, and axion-like particles. 
However, recent pathfinder experiments in the forward region of the LHC have successfully detected collider neutrinos for the first time~\cite{FASER:2023zcr,SNDLHC:2023pun} and demonstrated the potential to uncover new physics~\cite{FASER:2018ceo,Batell:2021blf,Foroughi-Abari:2020qar}. The FPF will offer a versatile range of experiments to take advantage of these physics opportunities by detecting neutrino interactions at the highest accelerator energies, thereby advancing our knowledge of particle interactions in the far-forward region ($\eta>7$).

The proposed site for the FPF is located 88\,m underground along the line-of-sight (LOS) of the ATLAS collision axis, approximately 627\,m west of the interaction point and shielded by over 200\,m of rock, as shown in \cref{fig:ExecutiveSummaryMap}. The facility will span roughly 75\,m in length and 11.8\,m in internal width, providing the necessary infrastructure to host a diverse range of experiments aimed at investigating physics phenomena at pseudorapidities above \( \eta \sim 7 \). An extensive site selection study has been conducted by the CERN Civil Engineering group~\cite{Boyd:2851822,PBCnote,PBCnote2,vibration-note}. Vibration, radiation, and safety studies have shown that the FPF can be constructed independently of the LHC without interfering with LHC operations. A core sample, taken along the location of the 88\,m deep shaft to provide information about the geological conditions, has confirmed that the site is suitable for construction. Studies of LHC-generated radiation have concluded that the facility can be safely accessed with appropriate controls during beam operations. 

\subsection{Experiments}

The FPF is uniquely suited to explore physics in the forward region because it will house a diverse set of experiments based on different detector technologies and optimized for particular physics goals. The proposed experiments are shown in \cref{fig:ExecutiveSummaryMap} and include~\cite{Feng:2022inv, Adhikary:2024nlv}:
%\vspace*{-0.1in}
\begin{itemize}
\setlength\itemsep{-0.01in}
\item FASER2, a magnetic tracking spectrometer, designed to search for light and weakly-interacting states, including new force carriers, sterile neutrinos, axion-like particles, and dark sector particles, and to distinguish $\nu$ and $\bar{\nu}$ charged current scattering.% in the upstream detectors. 
\item FASER$\nu$2, an on-axis emulsion detector, with pseudorapidity range $\eta > 8.4$, that will detect $\sim 10^6$ neutrinos at TeV energies with unparalleled spatial resolution, including several thousands of tau neutrinos, among the least studied of all the known particles.
\item FLArE, a 10-ton-scale, noble liquid, fine-grained time projection chamber that will detect neutrinos and search for light dark matter with high kinematic resolution, wide dynamic range and good particle-identification capabilities. 
\item FORMOSA, a detector composed of scintillating bars, with world-leading sensitivity to millicharged particles across a large range of masses.
\end{itemize}

\vspace{0.5em}

%------------------------
\begin{wrapfigure}{r}{0.5\textwidth}
    \vspace{-1.5em}
    \includegraphics[width=0.5\textwidth]{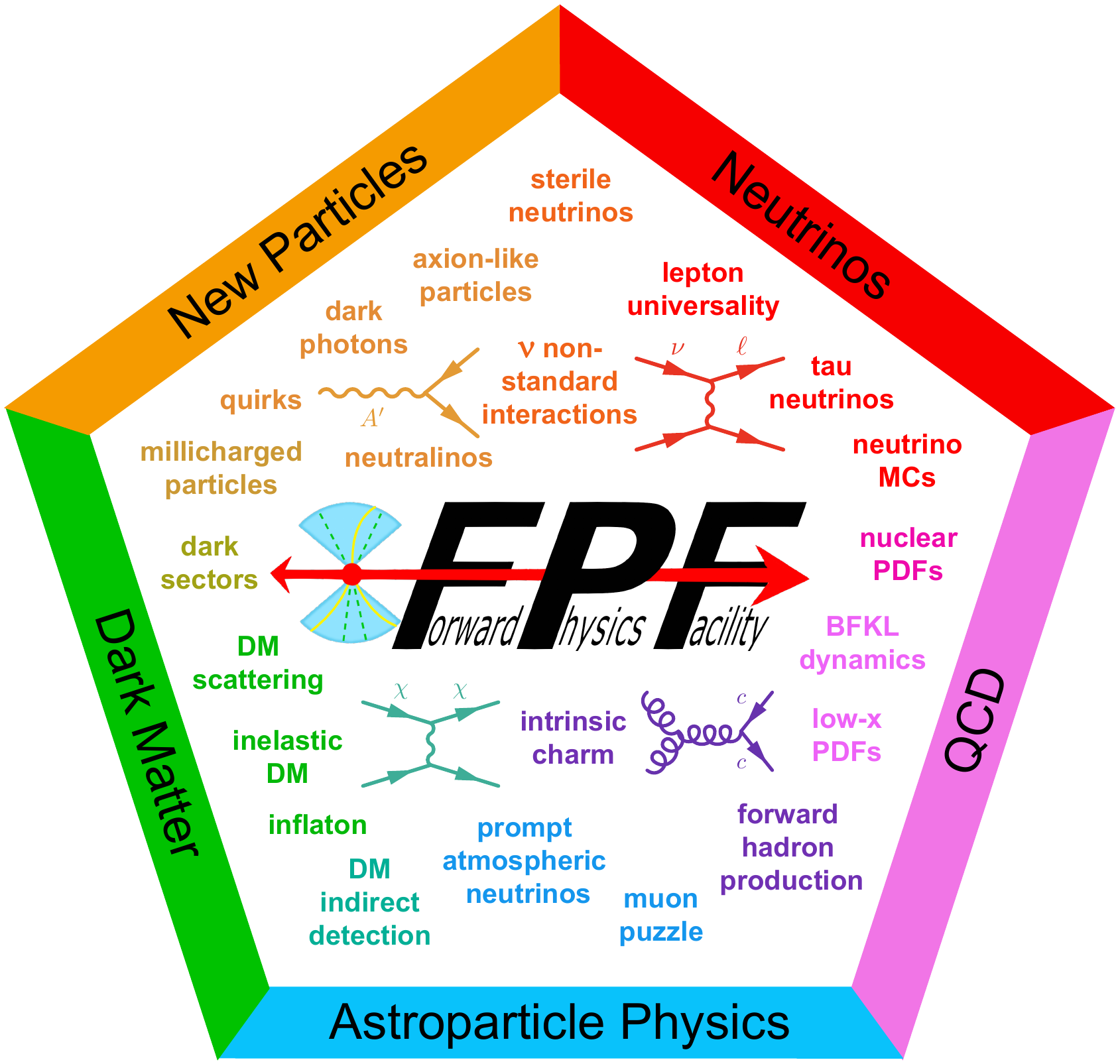}
    \vspace{-1.5em}
    \caption{\label{fig:PhysicsProgram}The rich physics program at the FPF spans many topics and frontiers~\cite{Feng:2022inv}.}
    \vspace{-2.em}
\end{wrapfigure}
%------------------------

As illustrated in \cref{fig:PhysicsProgram}, the FPF physics program encompasses a broad set of searches for novel new physics and unique Standard Model (SM) measurements that leverage the diverse capabilities of the suite of FPF experiments. The SM measurements specifically leverage the unprecedented flux of high-energy collider neutrinos that can be observed by FLArE and FASER$\nu$2 to study various physics scenarios, including strong connections to open questions in astroparticle physics. In the following, the synergies between the FPF physics program and current activities in astroparticle physics will be highlighted. A very comprehensive overview of the broad physics program of the FPF  can be found in Refs.~\cite{Anchordoqui:2021ghd,Feng:2022inv}.

%%% SECTION 3 %%%
\section{Astroparticle Physics at the FPF}
\label{sec:muon_puzzle}

The LHC neutrino fluxes depend sensitively on the mechanisms for forward light and heavy hadron production in $pp$ collisions~\cite{Kling:2021gos,Buonocore:2023kna}. The overall normalization of the muon-neutrino flux can be measured at the per-mille level at the FPF~\cite{Cruz-Martinez:2023sdv}. Thereby, the FPF provides unique opportunities for interdisciplinary studies at the intersection between high-energy particle and astroparticle physics~\cite{Anchordoqui:2021ghd,Feng:2022inv,Adhikary:2024nlv,Soldin:2023gox,Soldin:2024iev,Anchordoqui:2022ivb}, as outlined in the following.

\subsection{Light Hadron Production}
\label{sec:muon_puzzle}

Muons act as tracers of hadronic interactions, making their measurement in extensive air showers (EASs) essential for testing hadronic interaction models. For many years, several EAS experiments have reported discrepancies between model predictions and experimental data, a phenomenon known as the muon puzzle in EASs. Specifically, analyses from the Pierre Auger Observatory have shown a deficit in the number of muons compared to simulations~\cite{PierreAuger:2014ucz, PierreAuger:2016nfk,PierreAuger:2024neu}, and a systematic meta-analysis of data from nine different air shower experiments has revealed an energy-dependent trend in these discrepancies, with high statistical significance~\cite{Dembinski:2019uta,Soldin:2021wyv,Cazon:2020zhx}. Recent studies suggest that these discrepancies may be complex and indicate severe deficits in our understanding of particle physics~\cite{ArteagaVelazquez:2023fda}, which remain unresolved.

%------------------------
\begin{figure}[!b]
    \centering
    \captionsetup{width=1.\linewidth}
    \vspace{-1em}
    \includegraphics[width=1.\textwidth]{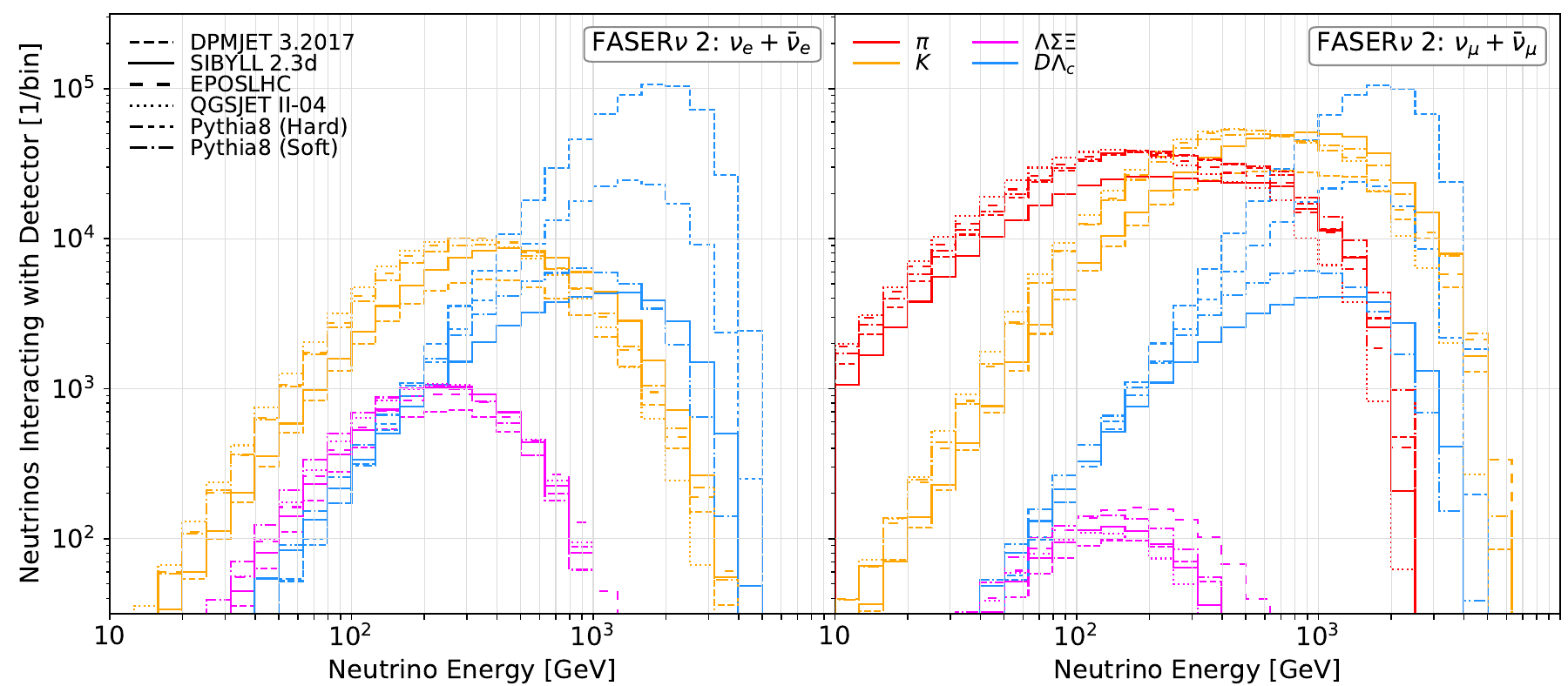}
    \caption{ \label{fig:AstroNuRate}Neutrino energy spectra\,\protect\cite{Feng:2022inv,Kling:2021gos} for electron neutrinos (left) and muon neutrinos (right) passing through FASER$\nu$2 for an integrated luminosity of $3\,\rm{ab}^{-1}$. The different production modes are shown separately, i.e., pion decays (red), kaon decays (orange), hyperon decays (magenta), and charm decays (blue). The predictions are obtained from SIBYLL-2.3d (solid), DPMJET-III.2017 (short dashed), EPOS-LHC (long dashed), QGSJET-II.04 (dotted), and Pythia~8.2 using soft-QCD processes (dot-dashed) and with hard-QCD processes for charm production (double-dot-dashed).}
    \vspace{-.5em}
\end{figure}
%------------------------

Dedicated measurements at the FPF will provide key insights into light hadron production in the far-forward region. The ratio of electron neutrino ($\nu_e$) to muon neutrino ($\nu_\mu$) fluxes, measured in FPF experiments, offers an indirect method for determining the ratio of charged kaons ($K$) to pions ($\pi$). Electron neutrinos are predominantly produced from kaon decays, while muon neutrinos originate from both pion and kaon decays. These neutrinos have distinct energy spectra, which allows them to be differentiated. Additionally, neutrinos from pion decays are more concentrated around the LOS compared to those from kaon decays, due to the lower mass of the pion ($m_\pi < m_K$), meaning pions impart less transverse momentum. As a result, the proximity of the neutrinos to the LOS, or their rapidity distribution, can help disentangle the origins of the neutrinos and provide an estimate of the pion-to-kaon ratio. 

\Cref{fig:AstroNuRate} shows predictions of the neutrino energy spectra for $\nu_e$ and $\nu_\mu$ interactions in FASER$\nu$2, assuming an integrated luminosity of $3\,\mathrm{ab}^{-1}$. These predictions, based on various models including SIBYLL-2.3d~\cite{Riehn:2019jet,Fedynitch:2018cbl,Riehn:2024prp}, DPMJET-III~\cite{Roesler:2000he}, EPOS-LHC~\cite{Pierog:2013ria}, QGSJET-II-04~\cite{Ostapchenko:2013pia,Ostapchenko:2005nj}, and Pythia 8~\cite{Sjostrand:2014zea,Fieg:2023kld}, show flux differences exceeding a factor of two, much larger than the expected statistical uncertainties at the FPF~\cite{Kling:2021gos,Cruz-Martinez:2023sdv}. Since the muon puzzle is thought to have its origins in soft-QCD processes~\cite{Albrecht:2021cxw}, this directly ties the FPF measurements to its QCD program. Dedicated QCD measurements at the FPF will therefore be crucial for improving our understanding of particle production in EASs.

\begin{figure}[tb]
    \centering
    \captionsetup{width=1.\linewidth}
    \vspace{-1em}
    \includegraphics[width=1.\textwidth]{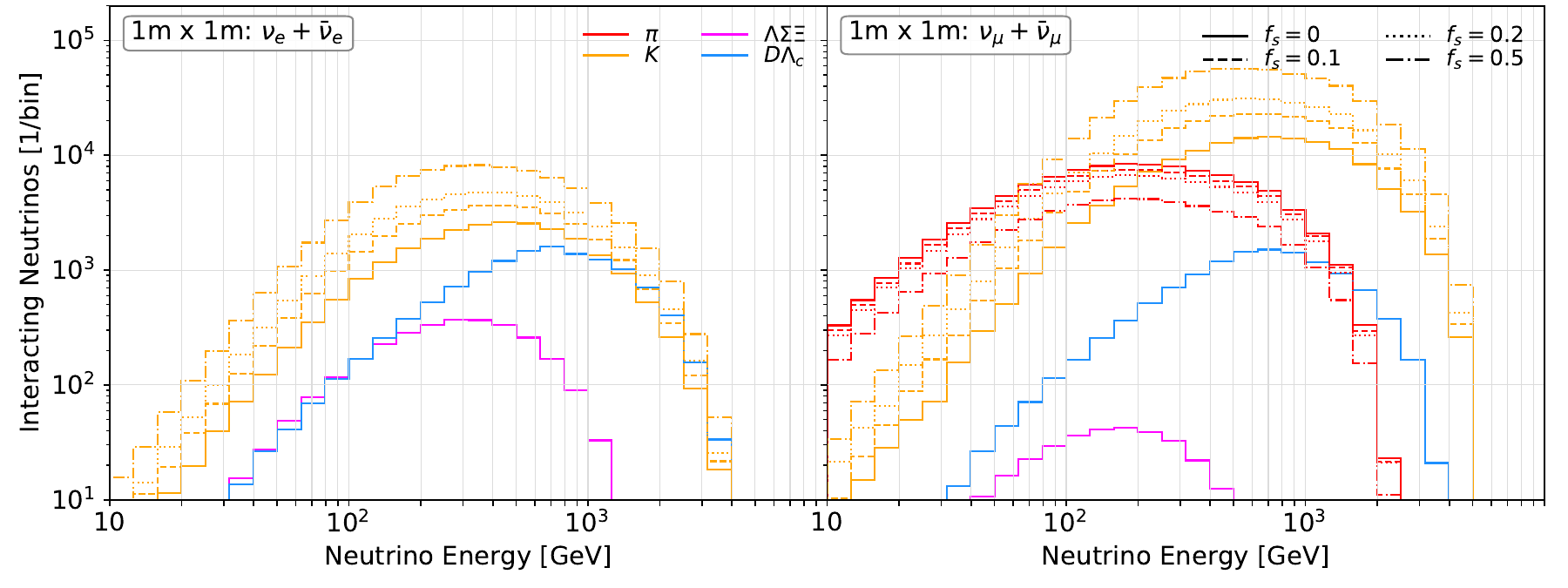}
    \vspace{-1.2em}
    \caption{\label{fig:NuSpectraFs}Neutrino energy spectra\,\protect\cite{Anchordoqui:2022fpn,Sciutto:2023zuz} for electron neutrinos (left) and muon neutrinos (right) passing through the FLArE detector. The vertical axis shows the number of neutrinos that pass the detector's cross-sectional area of $1\,\mathrm{m}^2$ for an integrated luminosity of $3\,{\rm ab}^{-1}$: pion decays (red), kaon decays (orange), hyperon decays (magenta), and charm decays (blue). The different line styles correspond to predictions obtained from SIBYLL-2.3d by varying $f_s$.}
\end{figure}

A potential key to resolving the muon puzzle may lie in the universal strangeness and baryon enhancement observed by the ALICE experiment in mid-rapidity collisions~\cite{ALICE:2016fzo}. This enhancement, which depends solely on particle multiplicity and not on specific details of the collision system, could allow for predictions of hadron composition in EASs in a phase space beyond current collider capabilities. If this enhancement increases in the forward region, it would affect muon production in EASs and could be traced by measuring the ratio of charged kaons to pions at the FPF. A specific example accounting for enhanced strangeness production is shown in \cref{fig:NuSpectraFs} where strangeness enhancement is introduced by allowing the substitution of pions with kaons in SIBYLL-2.3d with a probability $f_s$ at large pseudorapidities ($\eta > 4$)~\cite{Anchordoqui:2022fpn,Sciutto:2023zuz}. For $f_s = 0.1$ ($f_s = 0.2$), the predicted electron neutrino flux is $1.6$ ($2.2$) times higher at its peak than the baseline prediction. These differences are much larger than the expected uncertainties at the FPF. For values of $f_s$ between $0.4$ and $0.6$, this simple model can partially explain the observed data, suggesting a potential solution to the muon puzzle. This example demonstrates how the FPF will be uniquely positioned to test and constrain hadronic interaction models, providing a significant advancement in our understanding of multi-particle production in EASs.

%------------------------
%------------------------

\subsection{Charm Hadron Production}
\label{sec:neutrinos}

High-energy neutrinos of astrophysical origin are routinely observed by large-scale neutrino telescopes, such as IceCube~\cite{IceCube:2016zyt} and KM3NeT~\cite{Coniglione:2015aqa}, and atmospheric neutrinos produced in extensive air showers are an irreducible background for these searches. Neutrinos at high energies above 1~TeV are mainly produced in charm hadron decays. The production of charm quarks is dominated by gluon fusion and can be described using perturbative QCD~\cite{Gauld:2015kvh}. Measurements of the neutrino flux at the FPF therefore provide access to both the very high-$x$ and the very low-$x$ regions of the colliding protons. These measurements yield information about high-$x$ parton distribution functions (PDFs), in particular intrinsic charm, as well as novel QCD production mechanisms, such as BFKL effects and non-linear dynamics~\cite{Duwentaster:2023mbk}, well beyond the coverage of any existing experiments and providing key inputs for astroparticle physics, particularly for large-scale neutrino telescopes. 

One specific example of novel QCD dynamics that may affect the prompt neutrino fluxes is gluon saturation, which causes a suppression of the gluon density at low-$x$. This mechanism leads to a reduced flux of TeV-energy neutrinos at the FPF, as illustrated by the orange curve in \cref{fig:NuSpectraFs} (left). By defining tailored observables where theory uncertainties cancel out, such as the ratio between electron and tau neutrino event rates, FPF measurements can be used to measure the gluon PDF down to $x\sim 10^{-7}$~\cite{Rojo:2024tho}. Such measurements inform the study of novel QCD dynamics at small-$x$, a region where non-linear and BFKL-like effects are expected to dominate, as highlighted by the DGHP24 predictions~\cite{Duwentaster:2023mbk} for the gluon PDF based on saturation (recombination) effects. Thus, FPF measurements will provide stringent constraints on the  prompt atmospheric neutrino flux, contributing to the scientific program of large-scale neutrino telescopes. This is further quantified in \cref{fig:NuSpectraFs} (right), showing theoretical predictions for the prompt muon-neutrino flux based on the formalism of Refs.~\cite{Bai:2022xad,Bai:2021ira}, considering only PDF uncertainties (error bands), before and after FPF constraints are included. Although other sources of theory uncertainty also contribute to the total uncertainties, \cref{fig:NuSpectraFs} demonstrates the strong sensitivity of the FPF to the mechanisms governing atmospheric neutrino production from charm decays.

The robust interpretation of FPF measurements also requires state-of-the-art Monte Carlo event generators for neutrino scattering at TeV energies. Such generators, based on higher-order QCD corrections and matched to modern parton showers, are also relevant to model high-energy neutrino scattering at large-scale neutrino telescopes. Testing and validating neutrino event generators, such as the POWHEG-based ones presented in Refs.~\cite{vanBeekveld:2024ziz,FerrarioRavasio:2024kem,Buonocore:2024pdv}, with FPF data is also instrumental for the FPF beyond SM physics program, with neutrino signals representing the leading background in many such searches. 

%The LHC neutrino fluxes depend sensitively on the mechanisms for forward light and heavy hadron production in $pp$ collisions~\cite{Kling:2021gos,Buonocore:2023kna}. The overall normalization of the muon-neutrino flux can be measured at the per-mille level at the FPF~\cite{Cruz-Martinez:2023sdv}. Measurements of forward $D$-meson production at LHCb can constrain the gluon PDF down to $x\sim 10^{-5}$~\cite{Gauld:2016kpd,Zenaiev:2019ktw}. By defining tailored observables where theory uncertainties cancel out, such as the ratio between electron and tau neutrino event rates, FPF measurements can be used to pin down the gluon PDF down to $x\sim 10^{-7}$~\cite{Rojo:2024tho}, as shown in Fig.~\ref{fig:smallxQCD}. Such measurements inform the study of novel QCD dynamics at small-$x$, a region where non-linear and BFKL-like effects are expected to dominate, as highlighted by the DGHP24 predictions~\cite{Duwentaster:2023mbk} for the gluon PDF based on saturation (recombination) effects built into the DGLAP evolution. Constraints on the small-$x$ gluon PDF would be instrumental to inform FCC-hh cross sections, since at $\sqrt{s}=100$ TeV even Higgs and gauge boson production becomes a ``small-$x$'' process with potentially large corrections from BFKL resummation~\cite{Bonvini:2018vzv,Rojo:2016kwu}. These constraints on small-$x$ QCD are also relevant for astroparticle physics. \medskip

\begin{figure}[tb]
    \centering
    \captionsetup{width=1.\linewidth}
    \vspace{-1em}
    \includegraphics[width=.42\textwidth]{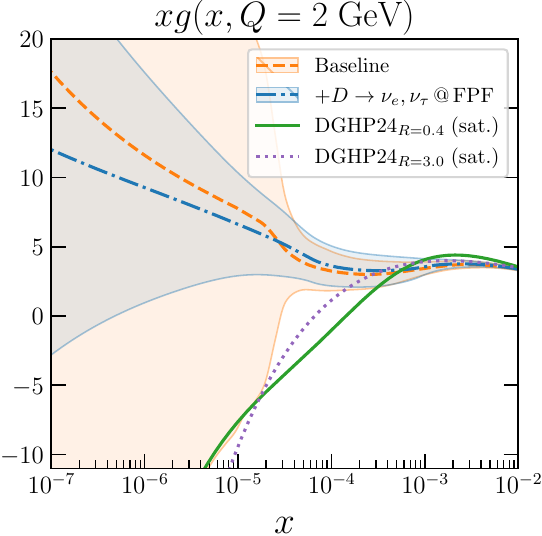}\;\;
    \includegraphics[width=.54\textwidth]{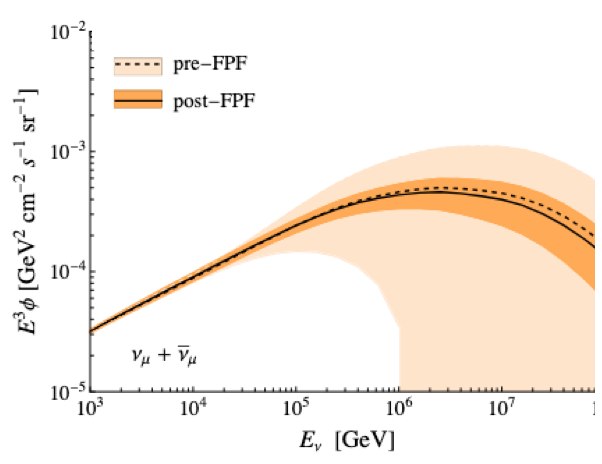}
    \caption{\label{fig:smallxQCD}Impact of FPF data on the small-$x$ gluon PDF (left)~\cite{Rojo:2024tho}, compared with non-linear QCD (saturation) models, and corresponding impact on the uncertainty of atmospheric muon-neutrino flux predictions (right)~\cite{Adhikary:2024nlv}.}
\end{figure}

%%% SECTION 4 %%%
\section{Conclusions}
\label{sec:conclusions}

The Forward Physics Facility is a proposal to establish a new underground laboratory at the high-luminosity LHC, dedicated to a range of experiments that will focus on particle production in the far-forward region. These experiments will enable groundbreaking neutrino measurements with unprecedented statistics, exploring hadron production in a phase space that is currently inaccessible to any existing collider experiment, but essential for accurately modeling high-energy hadronic interactions in the atmosphere. As a result, the measurements at the FPF are closely tied to key questions in astroparticle physics. They will provide crucial tests for constraining hadronic interaction models, leading to a deeper understanding of multi-particle production in extensive air showers and offering valuable insights into the muon puzzle. This will significantly reduce the uncertainties associated with ground-based cosmic-ray observations, which are often dependent on interpretations based on these models. Furthermore, the FPF will yield important data on charm hadron production in the forward region, enhancing predictions of the prompt atmospheric neutrino flux and reducing associated uncertainties in searches for astrophysical neutrinos. Thereby, the experiments at the FPF will play a pivotal role in advancing our understanding of the origin and nature of the highest-energy cosmic rays and astrophysical neutrinos, particularly in the context of future multi-messenger astrophysical observations~\cite{Schroder:2019rxp,Coleman:2022abf}.

%The Forward Physics Facility is a proposal to construct a new underground cavern at the high- luminosity LHC, which will host a variety of experiments focused on particle production in the far-forward region. These experiments will enable unique neutrino measurements with un- precedented statistics, probing hadron production in a phase space that is inaccessible to any existing collider experiment but crucial for modeling high-energy hadronic interactions in the atmosphere. Consequently, the measurements at the FPF are closely linked to open questions in modern astroparticle physics. They will provide unique tests to constrain hadronic interaction models, fundamentally improving our understanding of multi-particle production in extensive air showers, and thereby notably contribute to solving the muon puzzle. This will significantly reduce the uncertainties in ground-based cosmic-ray observations which typically rely on inter- pretations based on these models. Additionally, measurements at the FPF will provide crucial information about charm hadron production in the forward region which will improve predic- tions of the prompt atmospheric neutrino flux and reduce associated uncertainties in searches for astrophysical neutrinos. Thus, the experiments at the FPF will contribute to enhance our understanding of the origin and nature of the highest-energy cosmic rays and astrophysical neutrinos in the context of future multi-messenger observations~\cite{Schroder:2019rxp,Coleman:2022abf}.

\section*{Acknowledgements}
The author gratefully thanks the Forward Physics Facilty Initiative for useful discussions and input to this manuscript, in particular, L.~A.~Anchordoqui, J.~L.~Feng, F.~Kling, J.~Rojo, A.~M.~Stasto, M.~H.~Reno, and J.~Boyd.

%%% BIBLIOGRAPHY %%%
\bibliography{SciPost_BiBTeX_File.bib}

\end{document}